\documentclass[10pt,twocolumn,letterpaper]{article}

\usepackage[pagenumbers]{wacv} % To force page numbers, e.g. for an arXiv version

\usepackage[ruled]{algorithm2e} % For algorithms
\usepackage[utf8]{inputenc} % allow utf-8 input
\usepackage[T1]{fontenc}    % use 8-bit T1 fonts
\usepackage{url}            % simple URL typesetting
\usepackage{booktabs}       % professional-quality tables
\usepackage{amsfonts}       % blackboard math symbols
\usepackage{nicefrac}       % compact symbols for 1/2, etc.
\usepackage{microtype}      % microtypography
\usepackage{xcolor}         % colors
\usepackage{soul}
\usepackage{graphicx}
\usepackage{amsmath}
\usepackage{floatpag}
\usepackage{enumitem}
\usepackage{svg}
\usepackage{caption}
\usepackage{multirow}

\usepackage{amssymb}      % for \checkmark
\usepackage[table]{xcolor} %  for \rowcolor
\usepackage{arydshln}     % for \hdashline

\definecolor{wacvblue}{rgb}{0.21,0.49,0.74}
\usepackage[pagebackref,breaklinks,colorlinks,allcolors=wacvblue]{hyperref}

\def\wacvPaperID{2474} % *** Enter the WACV Paper ID here
\def\confName{WACV}
\def\confYear{2026}

\title{CAMP-VQA: Caption-Embedded Multimodal Perception for No-Reference Quality Assessment of Compressed Video}

\author{Xinyi Wang\qquad
Angeliki Katsenou\qquad
Junxiao Shen\qquad
David Bull\\
School of Computer Science, University of Bristol\\
Bristol, United Kingdom\\
{\tt\small \{xinyi.wang, angeliki.katsenou, junxiao.shen, dave.bull\}@bristol.ac.uk}
}
\begin{document}
\maketitle

\begin{abstract}
The prevalence of user-generated content (UGC) on platforms such as YouTube and TikTok has rendered no-reference (NR) perceptual video quality assessment (VQA) vital for optimizing video delivery. Nonetheless, the characteristics of non-professional acquisition and the subsequent transcoding of UGC video on sharing platforms present significant challenges for NR-VQA. Although NR-VQA models attempt to infer mean opinion scores (MOS), their modeling of subjective scores for compressed content remains limited due to the absence of fine-grained perceptual annotations of artifact types. To address these challenges, we propose CAMP-VQA, a novel NR-VQA framework that exploits the semantic understanding capabilities of large vision–language models. Our approach introduces a quality-aware prompting mechanism that integrates video metadata (e.g., resolution, frame rate, bitrate) with key fragments extracted from inter-frame variations to guide the BLIP-2 pretraining approach in generating fine-grained quality captions. A unified architecture has been designed to model perceptual quality across three dimensions: semantic alignment, temporal characteristics, and spatial characteristics. These multimodal features are extracted and fused, then regressed to video quality scores. Extensive experiments on a wide variety of UGC datasets demonstrate that our model consistently outperforms existing NR-VQA methods, achieving improved accuracy without the need for costly manual fine-grained annotations. Our method achieves the best performance in terms of average rank and linear correlation (SRCC: 0.928, PLCC: 0.938) compared to state-of-the-art methods. The source code and trained models, along with a user-friendly demo, are available at: {\footnotesize \url{https://github.com/xinyiW915/CAMP-VQA}}.
\end{abstract}

\begin{figure*}[t]
  \centering
  \includegraphics[width=0.915\textwidth]{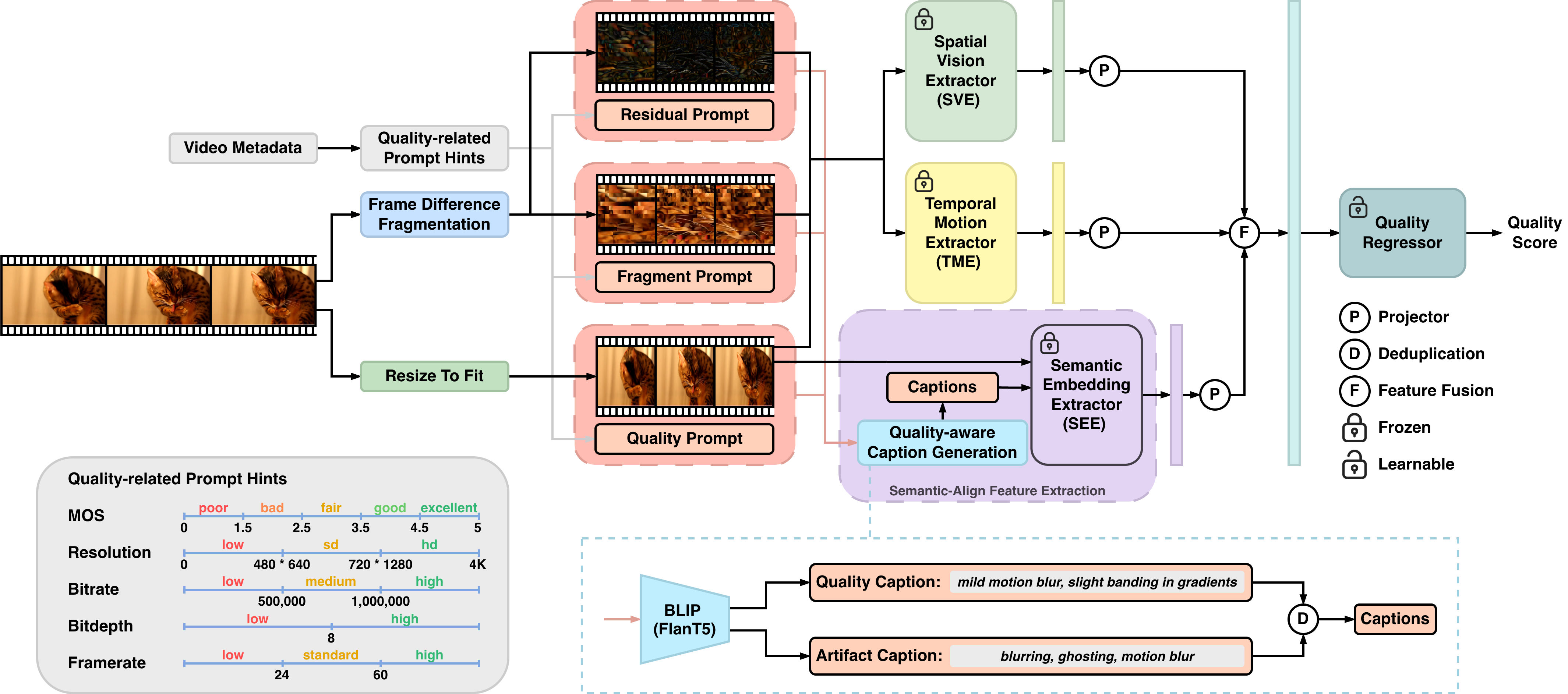}
  \caption{CAMP-VQA automatically generates quality-aware captions and fuses spatial, temporal, and semantic features to predict video quality.}
  \label{fig: camp-vqa}
\vspace{-2em}
\end{figure*}

\section{Introduction}
\label{sec:intro}
Recent advances in video compression technology have enabled the widespread use of streaming platforms such as YouTube and TikTok, over which billions of people actively exchange their User-Generated (Video) Content (UGC). Due to the impact of widely varying video acquisition conditions~\cite{lu2024kvq}, non-professional production environments, and bandwidth constraints, UGC videos often suffer a range of degradations, including noise, motion blur, out-of-focus blur, banding, and more~\cite{duan2025finevq}, leading to significant variations in visual quality that can severely degrade the user experience. Reliable video quality assessment techniques are therefore essential to ensure optimized streaming, leading to the best user satisfaction.

Early video quality assessment (VQA) methods primarily relied on mathematical expressions of pixel values between the original and compressed version (e.g., peak signal-to-noise ratio (PSNR) and structural similarity index (SSIM)~\cite{wang2004image}). These traditional full-reference VQA (FR-VQA) methods are still widely used because of their low complexity; they do not however effectively model human perception. While recent perceptually-aligned FR-VQA approaches based on machine learning, such as Video Multi-Method Assessment Fusion (VMAF)~\cite{li2016toward}, have achieved notable progress and adoption, they typically rely on the availability of pristine reference videos, which are unavailable in UGC scenarios. In contrast, no-reference VQA (NR-VQA) does not require the original reference video, so is of greater practical relevance in UGC scenarios. 

Traditional NR-VQA metrics incorporate hand-crafted features tailored to specific distortions~\cite{mittal2011blind, korhonen2019two, tu2021ugc} and perform well on small-scale datasets. Nevertheless, their accuracy drops when employed to assess the complex and varied spatio-temporal distortions associated with large-scale UGC content. Deep neural networks (DNNs) generalize better by leveraging their robust representation learning capabilities~\cite{min2024perceptual}. 

Perceptual NR-VQA typically follows a standard workflow: a sophisticated visual feature extractor aligned with human perception feeds features into a prediction module, such as regression or classification, to produce video quality scores. Therefore, model performance is largely dependent on the effectiveness of feature extraction. Currently, NR-VQA constructs perceptual feature extractors using neural networks pre-trained on large-scale image datasets. These extractors typically employ 2D-CNNs~\cite{li2019quality, tu2021rapique, li2021unified, madhusudana2023conviqt} or 3D-CNNs~\cite{liu2018end, sun2022deep, li2022blindly} and/or Transformers~\cite{you2021long, 10054487, wang2024relax} to capture perceptual features from sampled frames or fragments, effectively encoding visual and motion perception to analyze pixel variations caused by quality degradations. 
% The variation in resolution among UGC videos introduces a tradeoff between accuracy and computational efficiency during processing~\cite{zheng2024video}. Retaining the original resolution preserves the global visual information but incurs high computational costs; conversely, downscaling or cropping improves efficiency but may result in the loss of critical details. Given that deep neural networks require fixed input sizes, scaling can obscure local information, and conventional cropping often fails to reflect overall video quality. To address this, researchers have proposed fragment-based sampling strategies~\cite{wu2023neighbourhood, wu2023exploring, zhao2023zoom, wang2024relax, liu2024scaling} that capture local details while preserving global perceptual information, thus achieving a more optimal balance between efficiency and the accuracy of complex spatio-temporal feature extraction.

For all NR-VQA models, ground truth is based on subjective human ratings, such as mean opinion scores (MOS), which serve as supervisory signals and form the basis for modeling human perception of quality~\cite{mittal2011blind, korhonen2019two, tu2021ugc}. However, as a global quality label, MOS lacks explainability, and is thus not capable of guiding models to differentiate between specific types of distortion and semantic-level quality factors within videos. Building models with stronger artifact-semantic awareness requires incorporating more fine-grained annotations, such as blur, color bleeding, banding artifacts, etc~\cite{jia2024vqa, duan2025finevq}. Constructing such detailed annotation datasets is costly and time-consuming. As a response, recent studies have explored the use of multiple modalities. A prominent example is CLIP~\cite{radford2021learning}, which was trained on large-scale image-text pairs, establishing semantic associations between textual and visual information, making it suitable for semantic-aware feature extraction~\cite{he2024cover, yuan2024ptm, lu2024kvq} and perception-aware prompt design~\cite{wu2023towards, wu2023q, ge2024lmm}. Meanwhile, large language models (LLMs) and large multimodal models (LMMs) (e.g., GPT~\cite{brown2020language} and BLIP-2~\cite{li2023blip}) demonstrate remarkable performance in semantic understanding, enabling prompt-driven quality assessment by leveraging multimodal features. However, training LMMs for VQA is heavily based on hand-crafted question–answer pairs designed for low-level vision tasks. Hence, a key challenge remains: to develop a unified framework capable of automatically extracting semantic-aware features and integrating them with spatio-temporal features to assess video quality under various distortions, without the need for detailed expert annotations.

% To reduce the laborious low-level vision tasks on fine-grained annotations and to fully leverage the semantic artifact understanding and multimodal representation capabilities of pre-trained large vision-language models (VLMs), this paper proposes \textit{a novel NR-VQA framework guided by quality-aware prompts: CAMP-VQA}. 
To avoid the laborious fine-grained annotations in low-level vision tasks and to capitalize on the semantic and multimodal capabilities of pre-trained vision-language models (VLMs), we propose \textit{CAMP-VQA, a novel NR-VQA framework guided by quality-aware prompts}.
CAMP-VQA introduces a quality-aware prompting mechanism incorporating video metadata, such as resolution, frame rate, and bitrate, alongside key fragments extracted from inter-frame variations, effectively highlighting distorted regions. Our method constructs quality-sensitive prompts to guide a vision-language model, BLIP-2~\cite{li2023blip}, in generating fine-grained quality captions aligned with human perception. Hence, we model deep video features across three dimensions: semantic, temporal, and spatial. The CAMP-VQA model, as illustrated in Fig.~\ref{fig: camp-vqa}, comprises three feature components: the Semantic artifact Embedding Extractor (SEE), which extracts semantic artifact embeddings aligned with visual information; the Temporal Motion Extractor (TME), built on SlowFast~\cite{feichtenhofer2019slowfast} as the backbone; and the Spatial Vision Extractor (SVE), built on Swin-Large~\cite{liu2021swin}. Moreover, the perceptual features of the three components are fused and fed into a multilayer perceptron (MLP) regressor to enable training and prediction of video quality scores. Our unified framework constructs perceptual-aware features by integrating semantic artifact perception with spatio-temporal information, thereby significantly enhancing the accuracy and perceptual capability of NR-VQA while mitigating the need for extensive, detailed human annotations.
% We conducted extensive experiments on multiple public UGC datasets, demonstrating that CAMP-VQA significantly enhances quality prediction accuracy without requiring additional fine-grained annotations.

Our main contributions can be summarized as follows:
\begin{enumerate}[leftmargin=*, itemsep=2pt, topsep=2pt, parsep=0pt]
    \item A quality-aware prompting mechanism is proposed to guide a pre-trained VLM in generating meaningful, quality-related captions by integrating video metadata with distortion-focused fragments extracted based on inter-frame variations.
    \item A unified multimodal NR-VQA framework is developed, which fuses deep features from semantic, temporal and spatial dimensions.
    \item Extensive validation of our model on multiple public UGC datasets representing a variety of UGC use cases, including different orientations, genres, short-form videos, and gaming. This provides evidence of CAMP-VQA's superior performance compared to existing NR-VQA models.
\end{enumerate}

\section{Related Works On NR-VQA}
\label{sec:bg}
\textbf{Hand-crafted NR-VQA Models:} Traditional NR-VQA models relied on image data statistics. For example, BRISQUE~\cite{mittal2011blind} performs well for certain types of distortions such as banding, noise, and blur. TLVQM~\cite{korhonen2019two} and VIDEVAL~\cite{tu2021ugc} integrate hand-crafted, quality-aware features with feature selection strategies and employ shallow Support Vector Regression (SVR) heads to predict quality scores, demonstrating good performance on standard video datasets. However, the limited representational capacity of hand-crafted features often limits their ability to capture complex factors affecting the quality of in-the-wild videos, thereby limiting their generalizability.

\noindent
\textbf{Deep NR-VQA Models:} Enabled by the semantic awareness of DNNs and the availability of large-scale UGC datasets, deep NR-VQA models have emerged as the mainstream approach. VSFA~\cite{li2019quality} employs GRUs to aggregate frame-level features extracted using ResNet-50~\cite{he2016deep}, effectively capturing motion variations that are often missed by frame-based methods. More recent developments have addressed the challenges posed by multi-resolution content and complex spatio-temporal distortions that are manifest in UGC. PVQ~\cite{ying2021patch} extracts both 2D and 3D features from cropped video patches to capture local and global perceptual quality. Li et al.~\cite{li2022blindly} utilize transfer learning from image quality assessment datasets for spatial feature extraction and apply a pre-trained action recognition DNN for motion perception. RankDVQA~\cite{feng2024rankdvqa} uses hybrid training by using VMAF as a proxy. Given the high computational cost of processing high-resolution UGC videos, recent methods have adopted fragment-based or sampling strategies to balance accuracy and efficiency. Fast-VQA~\cite{wu2023neighbourhood} and DOVER~\cite{wu2023exploring} employ a grid mini-patch sampling (GMS) strategy to extract fragments that preserve global video quality. Additionally, DOVER incorporates an aesthetic branch using inflated ConvNeXt~\cite{liu2022convnet} to reflect subjective preferences. Zoom-VQA~\cite{zhao2023zoom} presents a dual-branch perception network that independently processes image and clip inputs. SAMA~\cite{liu2024scaling} implements a scalable, masking-based video sampling approach to construct multi-scale features. ReLaX-VQA~\cite{wang2024relax} and DIVA-VQA~\cite{wang2025diva} intelligently select spatio-temporal fragments based on frame differences and further enhance feature abstraction through DNN layer-stacking techniques and spatio-temporal feature extraction, respectively.

\noindent
\textbf{Large Multimodal NR-VQA Models:} The rapid advancement of LLMs and LMMs has significantly enhanced the capabilities of low-level vision understanding methods, with NR-VQA gradually incorporating similar semantic feature modeling techniques. Methods such as MaxVQA~\cite{wu2023towards}, PTM-VQA~\cite{yuan2024ptm}, COVER~\cite{he2024cover}, and KSVQE~\cite{lu2024kvq} leverage the CLIP~\cite{radford2021learning} pre-trained vision encoder to model semantic-level features in videos, greatly improving performance in quality prediction under complex scenarios. FineVQ~\cite{duan2025finevq} further strengthens fine-grained quality perception for UGC videos, supporting quality rating, scoring, and attribution. Meanwhile, Q-Align~\cite{wu2023q} and LMM-VQA~\cite{ge2024lmm} fine-tune LMMs to effectively integrate visual and textual representations that closely align with human perception. These multimodal NR-VQA models rely on language-driven paradigms and require laborious fine-grained annotations by subjects, rendering this impractical. Our method effectively addresses this challenge by leveraging quality-aware caption generation, enabling the automatic derivation of fine-grained, quality-related captions in place of manual annotation.
% These multimodal approaches exhibit strong potential in capturing the complex characteristics of video quality across different scenarios, marking a shift from methods reliant on human subjective scoring towards language-prompt-driven modeling paradigms.

\section{Proposed Framework}
\label{sec:method}
The basic concepts and components of the CAMP-VQA framework (see Fig.~\ref{fig: camp-vqa}) are introduced in this section.

\subsection{Frame Difference Fragmentation} 
Since in compression, quality degradation mainly occurs in regions with visual changes (e.g., large translational or small irregular motion), we employ the frame difference fragmentation (FDF) module~\cite{wang2024relax, wang2025diva} to obtain frame fragments that highlight distortion-prone regions, and residual fragments that indicate motion changes within a video sequence. This approach enables the construction of more specific prompts that guide the VLM model to generate quality-aware descriptions rather than generic content summaries.

In particular, given the current frame $F_t$ and its predecessor $F_{t-1}$, the absolute difference of pixel $(i, j)$ produces the residual $R_t$:
% \begin{equation}
% \label{eqn:01}
$R_t(i,j) = |F_t(i,j) - F_{t-1}(i,j)|.$
% \end{equation}
$R_t$ is divided into a set of non-overlapping patch regions of size $p \times p$, indexed by $k$. The residual intensity (i.e., the degree of difference) of the $k$-th patch is defined as:
\begin{equation}
\label{eqn:02}
\resizebox{.8\hsize}{!}{$
\Delta_k = \sum_{x=i}^{p} \sum_{y=j}^{p} \left| \mathcal{P}_t^{(k)}(x,y) - \mathcal{P}_{t-1}^{(k)}(x,y) \right|,
$}
\end{equation}
where $\mathcal{P}_t^{(k)}(x, y)$ and $\mathcal{P}_{t-1}^{(k)}(x, y)$ denote the pixel values at position $(x, y)$ within the $k$-th patch of the current and previous frames, respectively. All patches are ranked according to their corresponding $\Delta_k$, and the top $K$ patches with the highest degrees of difference are selected as key regions within the current frame. These regions typically correspond to notable spatio-temporal variations. More details on FDF are provided in the \textit{Supplementary Material}.

\subsection{Semantic-aligned Feature Extraction}
\begin{figure}[htbp]
    \centering
    \includegraphics[width=1\linewidth]{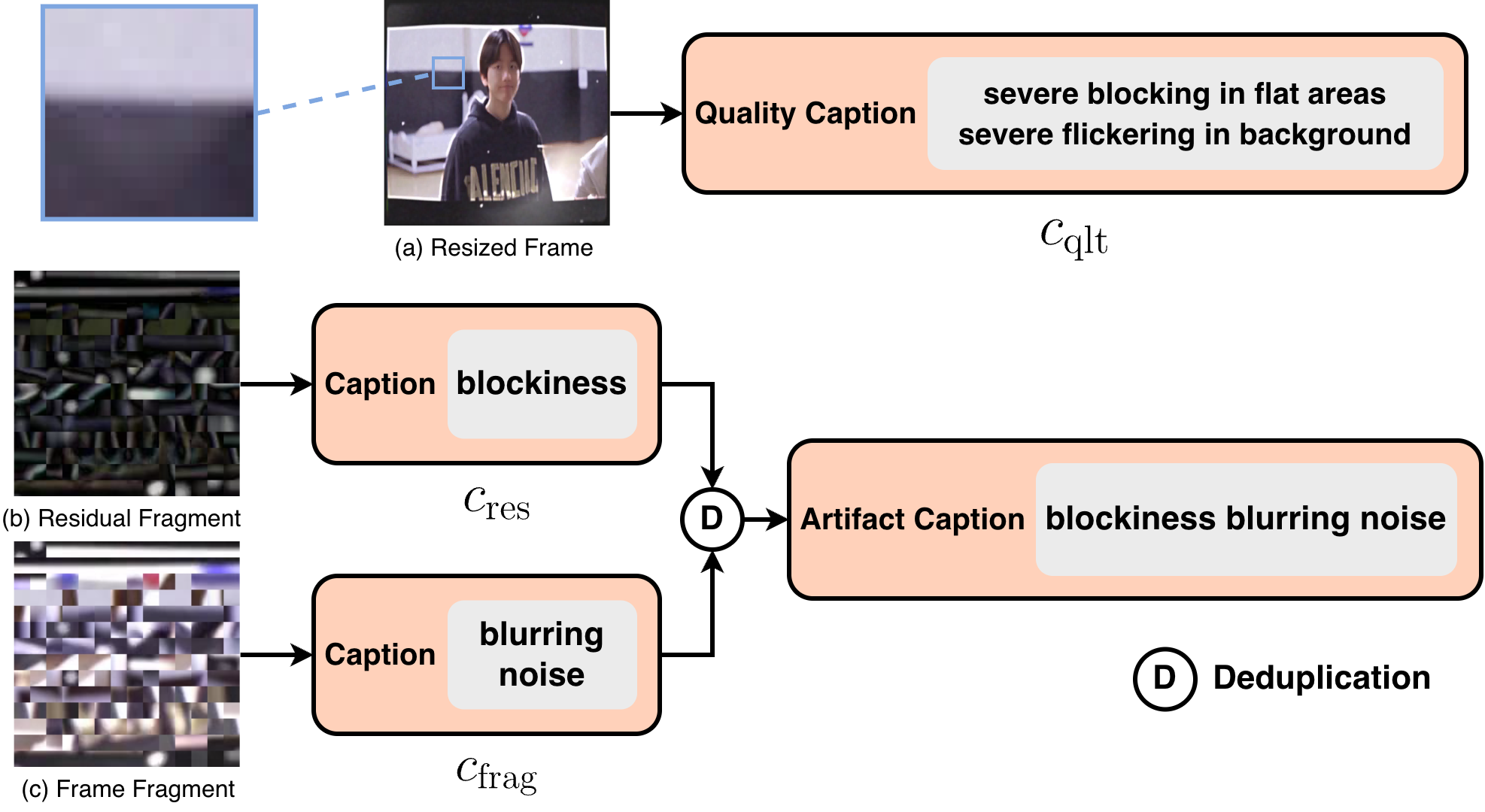}
    \caption{Example of quality-aware captioning from a sampled video frame and fragments.}
    \label{fig: SemanticFeature}
\end{figure}

\noindent
\textbf{Quality-aware Caption Generation:}
To address the scarcity of fine-grained manual quality annotations in low-level vision VQA datasets, we propose a quality-aware caption generation module guided by video metadata. This module automatically extracts quality and artifact captions from video frames to serve as fine-grained annotations. We employ the BLIP-2 VLM model~\cite{li2023blip}, which has demonstrated high levels of prompt responsiveness and cross-modal understanding. Guided by quality-related prompts, it generates descriptive texts that align more closely with human perception. Leveraging low-level image features, we use prompt engineering to generate quality-aware captions, which are then mapped into a unified vision-language embedding space.

Through our FDF module, two fragments are constructed for each sampled frame $F_t$: a residual fragment $\mathbf{F_t}^\text{res}$ and a frame fragment $\mathbf{F_t}^{\text{frag}}$, which represent distortions and contextual variations. We design distinct base prompts for the above three types of image inputs, based on the quality attributes identified in subjective quality assessment~\cite{jia2024vqa}. Guided by this predefined set of prompts, the Flan-T5 decoder~\cite{chung2024scaling} generates three quality-aware captions ($c_{\text{qlt}}$ at the frame level, and $c_{\text{res}}$ and $c_{\text{frag}}$ at the fragment level), as illustrated in Fig.~\ref{fig: SemanticFeature}:
\begin{equation}
\label{eqn:03}
\resizebox{.91\hsize}{!}{$
c_{\text{qlt}} = \mathcal{G}(F_t,\mathcal{P}_{\text{qlt}}),\;
c_{\text{res}} = \mathcal{G}(\mathbf{F_t}^{\text{res}},\mathcal{P}_{\text{res}}),\;
c_{\text{frag}} = \mathcal{G}(\mathbf{F_t}^{\text{frag}},\mathcal{P}_{\text{frag}}),
$}
\end{equation}
where $\mathcal{G}(\cdot, \mathcal{P})$ denotes the vision-language generation process guided by the prompt $\mathcal{P}$. We also incorporate video metadata, such as resolution and bitrate, to dynamically adjust the prompts. We design differentiated quality hints based on different metadata ranges to guide the base prompt, thereby improving the alignment between the generated text and quality perception. For detailed prompt configurations and more caption generation examples, refer to the \textit{Supplementary Material}.

\noindent
\textbf{Semantic Embedding Extractor:} 
After data cleaning, the generated descriptions are categorized into two types of inputs: the \textit{quality caption}, derived from sampled frames, and the \textit{artifact captions}, generated from the two auxiliary fragments $\mathbf{F_t}^\text{res}$ and $\mathbf{F_t}^{\text{frag}}$. These textual inputs, along with the sampled frames, are fed into the SEE, which is based on the CLIP~\cite{radford2021learning} backbone, a contrastive vision-language model that embeds images and texts into a shared semantic space~\cite{ghildyal2025foundation}. Through the text and image encoders, three types of semantic embeddings are extracted: 
\begin{equation}
\label{eqn:04}
\mathbf{e}^{\text{img}} = \phi_{\text{i}}(F_t),\
\mathbf{e}^{\text{qlt}} = \phi_{\text{t}}(\mathbf{c}_{\text{qlt}}),\
\mathbf{e}^{\text{art}} = \phi_{\text{t}}(\mathbf{c}_{\text{res}} + \mathbf{c}_{\text{frag}}),
\end{equation}
where $\phi_i(\cdot)$ and $\phi_t(\cdot)$ denote the image and text encoder, respectively, and $+$ represents the concatenation and integration operation for artifact captions.

We apply global average pooling (GAP) to these three semantic embeddings extracted from the sampled frame sequences, thus generating global embeddings for the video to facilitate subsequent fusion:
\begin{equation}
\label{eqn:05}
\resizebox{.91\hsize}{!}{$
\bar{\mathbf{e}}^{(m)} = \text{GAP}(\{\mathbf{e}_t^{(m)}\}_{t=1}^{T}) \in \mathbb{R}^{d_m},\;
m \in \{\text{img}, \text{qlt}, \text{art}\},\;
d_m \in \{d_i, d_t, d_t\},
$}
\end{equation}
where $t$ denotes the temporal index of the sampled frames at half the frame rate, $m$ denotes the embedding type with $\texttt{img}$ of the frames, $\texttt{qlt}$ of the quality captions, $\texttt{art}$ of the artifact captions, and $d_m$ is the dimensionality of each embedding vector, with $d_i$ and $d_t$ for the image and text embeddings respectively. The three resulting feature categories span the dimensions of image semantics, quality perception, and artifact perception. These representations establish a basis for the  subsequent multimodal feature fusion $\mathbf{z}_{\text{semantic}}$:
\begin{equation}
\label{eqn:06}
\mathbf{z}_{\text{semantic}} = [\bar{\mathbf{e}}^{\text{img}} \, \| \, \bar{\mathbf{e}}^{\text{qlt}} \, \| \, \bar{\mathbf{e}}^{\text{art}}] \in \mathbb{R}^{d_i + 2d_t}.
\end{equation}

\section{Spatio-temporal Feature Extraction}
We employ TME and SVE to extract spatio-temporal features. To model temporal dynamics, TME adopts a dual-pathway design based on a SlowFast~\cite{feichtenhofer2019slowfast} backbone to capture motion patterns at different temporal scales. GAP is applied to extract slow and fast motion features from each pathway, and temporal features are concatenated as follows:
\begin{equation}
\label{eqn:07}
\mathbf{z_{slowfast}} = \left[ \mathbf{z}_{\text{slow}} ; \mathbf{z}_{\text{fast}} \right] \in \mathbb{R}^{B \times (d_s + d_f)},
\end{equation}
where $\mathbf{z}_{\text{slow}}$ and $\mathbf{z}_{\text{fast}}$ denote pooled features of the slow and fast pathways, respectively, while $B$ is the batch size and $d_s$, $d_f$ are the corresponding feature dimensions.

For spatial representations, SVE employs SwinT~\cite{liu2021swin} encoder $\phi_\text{swin}$ to extract and capture long-range dependencies in images. GAP is applied to obtain spatial features:
\begin{equation}
\label{eqn:08}
z_{\text{swint}} = \text{GAP}(\phi_\text{swin}(\mathbf{X_\text{frame}})) \in \mathbb{R}^{N \times d_s},
\end{equation}
where $\mathbf{X_\text{frame}}$ denotes input frames of a video clip in tensor form, $N$ is the number of frames, and $d_s$ is the spatial feature dimension. Detailed formulations and explanations are provided in the \textit{Supplementary Material}.

\begin{table*}[t]
    \centering
    \footnotesize
    \setlength{\tabcolsep}{2.8pt}
    \caption{Performance comparison of the evaluated NR-VQA models on the six NR-VQA datasets. The \textcolor{red}{\textbf{red}} and \textcolor{blue}{\textbf{blue}} entries indicate the 1st and 2nd best on each database for each performance metric, respectively. Extra data refers to additional datasets that models use for training.}
    \vspace{-1em}
    \label{tab: ComparisonToSoA}
    \begin{tabular}{@{}llcccccccccccccc@{}}
    \toprule 
    \multicolumn{2}{l}{\textbf{Target Quality Dataset}} & \multicolumn{2}{c}{\textbf{CVD2014}} & \multicolumn{2}{c}{\textbf{KoNViD-1k}} & \multicolumn{2}{c}{\textbf{LIVE-VQC}} & \multicolumn{2}{c}{\textbf{YouTube-UGC}}  & \multicolumn{2}{c}{\textbf{LSVQ\(_{\text{test}}\)}} & \multicolumn{2}{c}{\textbf{LSVQ\(_{\text{1080p}}\)}} & \multicolumn{2}{c}{\textbf{FineVD}}\\
    
    \cmidrule(lr){1-2} \cmidrule(lr){3-4} \cmidrule(lr){5-6} \cmidrule(lr){7-8} \cmidrule(lr){9-10} \cmidrule(lr){11-12} \cmidrule(lr){13-14} \cmidrule(lr){15-16}
    Model & \shortstack{Extra\\Data} & SRCC & PLCC & SRCC & PLCC & SRCC & PLCC & SRCC & PLCC & SRCC & PLCC & SRCC & PLCC & SRCC & PLCC\\
    \midrule
    % BRISQUE\cite{mittal2011blind} &N/A &0.555 &0.553 &0.678 &0.675 &0.610 &0.665 &0.352 &0.377 &0.579 &0.576 &0.497 &0.531\\
    TLVQM\cite{korhonen2019two} &N/A &0.540 &0.579 &0.762 &0.746 &0.813 &0.791 &0.680 &0.688 &0.772 &0.774 &0.589 &0.616 &0.654 &0.655\\
    VIDEVAL\cite{tu2021ugc} &N/A &0.766 &0.806 &0.807 &0.792 &0.773 &0.775 &0.781 &0.793 &0.794 &0.783 &0.545 &0.554 &0.731 &0.731\\
    \hdashline
    \noalign{\vskip 2pt}
    VSFA\cite{li2019quality} &None &0.870 &0.868 &0.773 &0.775 &0.773 &0.795 &0.724 &0.743 &0.801 &0.796 &0.675 &0.704 &0.773 &0.793\\
    PVQ\cite{ying2021patch} &LSVQ &N/A &N/A &0.791 &0.786 &0.827 &0.837 &N/A &N/A &0.827 &0.828 &0.711 &0.739 &N/A &N/A\\
    BVQA\cite{li2022blindly} &None &0.872 &0.869 &0.834 &0.836 &0.834 &0.842 &0.818 &0.826 &0.852 &0.855 &0.771 &0.782 &N/A &N/A\\
    SimpleVQA\cite{sun2022deep} &None &N/A &N/A &0.856 &0.860 &0.845 &0.859 &0.847 &0.856 &0.867 &0.861 &0.764 &0.803 &0.831 & 0.836\\
    \hdashline
    \noalign{\vskip 2pt}
    FAST-VQA\cite{wu2023neighbourhood} &LSVQ &0.891 &0.903 &0.891 &0.892 &0.849 &0.862 &0.855 &0.852 &0.876 &0.877 &0.779 &0.814 &0.835 &0.847\\
    Zoom-VQA\cite{zhao2023zoom} &LSVQ &N/A &N/A &0.877 &0.875 &0.814 &0.833 &N/A &N/A &0.886 &0.879 &0.799 &0.819 &N/A &N/A\\
    DOVER\cite{wu2023exploring} &LSVQ &0.858 &0.881 & 0.909 & 0.906 & 0.860 &0.875 & 0.890 &0.891 &0.888 &0.889 &0.795 &0.830 &0.842 &0.839\\
    SAMA\cite{liu2024scaling} &LSVQ &N/A &N/A &0.892 &0.892 &0.860 &0.878 &0.881 &0.880 &0.883 &0.884 &0.782 &0.822 &N/A &N/A\\
    ReLaX-VQA\cite{wang2024relax} &LSVQ &0.897 &0.929 &0.872 &0.867 &0.847 &0.888 &0.847 &0.865 &0.869 &0.869 &0.768 &0.810 &N/A &N/A\\
    \hdashline
    \noalign{\vskip 2pt}
    PTM-VQA\cite{yuan2024ptm} &None &N/A &N/A &0.857 &0.872 &0.811 &0.820 &0.858 &0.857 &0.864 &0.855 &0.782 &0.736 &N/A &N/A\\
    COVER\cite{he2024cover} &None &N/A &N/A &0.893 &0.895 &0.809 &0.848 &\textcolor{red}{\textbf{0.914}} &0.917 &N/A &N/A &N/A &N/A &N/A &N/A\\
    KSVQE\cite{lu2024kvq} &KVQ &N/A &N/A &0.922 &0.921 &0.861 &0.883 &0.900 &0.912 &0.886 &0.888 &0.790 &0.823 &N/A &N/A\\
    LMM-VQA\cite{ge2024lmm} &LSVQ &N/A &N/A &\textcolor{blue}{\textbf{0.929}} &0.930 &0.891 &0.903  &0.901 &0.897 &\textcolor{blue}{\textbf{0.916}} &\textcolor{blue}{\textbf{0.919}} &\textcolor{blue}{\textbf{0.891}} &\textcolor{blue}{\textbf{0.899}} &N/A &N/A\\
    FineVQ\cite{duan2025finevq} &FineVD &N/A &N/A &0.915 &0.910 &0.895 &0.895 &0.910 &0.914 &0.900 &0.900 &0.828 &0.857 &0.883 &0.889\\
    \hdashline
    \noalign{\vskip 2pt}
    \rowcolor{gray!15}
    \textbf{CAMP-VQA} &None &\textcolor{blue}{\textbf{0.933}} &\textcolor{blue}{\textbf{0.944}} &0.927 &\textcolor{blue}{\textbf{0.936}} &\textcolor{blue}{\textbf{0.922}} &\textcolor{blue}{\textbf{0.940}} &0.901 &\textcolor{blue}{\textbf{0.920}} &\textcolor{red}{\textbf{0.920}} &\textcolor{red}{\textbf{0.933}} &\textcolor{red}{\textbf{0.908}} &\textcolor{red}{\textbf{0.920}} &\textcolor{blue}{\textbf{0.919}} &\textcolor{blue}{\textbf{0.923}}\\
    \rowcolor{gray!15}
    \textbf{CAMP-VQA (\textit{w/ fine-tune})} &LSVQ &\textcolor{red}{\textbf{0.966}} &\textcolor{red}{\textbf{0.964}} &\textcolor{red}{\textbf{0.930}} &\textcolor{red}{\textbf{0.944}} &\textcolor{red}{\textbf{0.934}} &\textcolor{red}{\textbf{0.946}} &\textcolor{blue}{\textbf{0.912}} &\textcolor{red}{\textbf{0.928}} &\textcolor{red}{\textbf{0.920\textsuperscript{*}}} &\textcolor{red}{\textbf{0.933\textsuperscript{*} }}&\textcolor{red}{\textbf{0.908\textsuperscript{*}}} &\textcolor{red}{\textbf{0.920\textsuperscript{*}}} &\textcolor{red}{\textbf{0.924}} &\textcolor{red}{\textbf{0.933}}\\
    \bottomrule
    \end{tabular}
    \vspace{-1em}
\end{table*}

\subsection{Quality Prediction}
\textbf{Multimodal Video Feature Fusion:} 
The subjective perception of video quality is influenced by multiple factors, including visual semantics, temporal dynamics, and spatial structure. To model these perceptual factors, we have designed a multimodal feature fusion module that integrates perceptual information from three modalities: semantic, temporal, and spatial. Specifically, given an input video $V$, we first divide it according to a frame rate $r$, producing a sequence of video segments of length $T$: 
$S_i = V[t_i : t_i + T], \quad t_i = i \cdot r,\ i = 0, 1, 2, \dots, M-1.$
If a segment contains fewer than $T$ frames, the last frame is replicated to ensure a uniform length. For each video segment $S_i$, we employ the FDF module to extract three key components that capture global and local visual information: the frame sequence $\mathbf{F}_{\text{frame}}$, the residual fragments $\mathbf{F}_{\text{resfrag}}$, and the frame fragments $\mathbf{F}_{\text{frag}}$:
\begin{equation}
\label{eqn:10}
[\mathbf{F}_{\text{frame}}^{(i)},\ \mathbf{F}_{\text{resfrag}}^{(i)},\ \mathbf{F}_{\text{frag}}^{(i)}] = \text{FDF}(S_i).
\end{equation}
These three components are jointly fed into the TME and the SVE to encode the corresponding temporal features and spatial features. For semantic modeling, we used BLIP-2 to generate quality-aware captions that reflect fine-grained perceptions related to video quality. Then, the sampled frame sequence $\mathbf{F}_{\text{sampled}}$ from each video segment, together with the corresponding quality-aware $\texttt{caption}$ (i.e., $c_{\text{qlt}}$, $c_{\text{res}}$, and $c_{\text{frag}}$), is fed into SEE to obtain quality-aware semantic embeddings:
\begin{equation}
\label{eqn:11}
\resizebox{.6\hsize}{!}{$
\begin{aligned}
\mathbf{f}_{\text{SE}}^{(i)} = \text{SEE}(\mathbf{F}_{\text{sampled}}^{(i)},\ \text{caption}^{(i)}) \\
\mathbf{f}_{\text{TM}}^{(i)} = \text{TME}([\mathbf{F}_{\text{frame}}^{(i)},\ \mathbf{F}_{\text{resfrag}}^{(i)},\ \mathbf{F}_{\text{frag}}^{(i)}]) \\
\mathbf{f}_{\text{SV}}^{(i)} = \text{SVE}([\mathbf{F}^{(i)}_{\text{frame}},\ \mathbf{F}^{(i)}_{\text{resfrag}},\ \mathbf{F}^{(i)}_{\text{frag}}]).
\end{aligned}
$}
\end{equation}
We used GAP to project the three modalities in each video segment. The pooled features are then averaged to obtain the global semantic, temporal, and spatial representation vectors for the video:
\begin{equation}
\label{eqn:12}
\mathbf{f}_c = \frac{1}{M} \sum_{i=0}^{M-1} \mathbf{f}^{(i)}_c, \quad c \in \{\text{SE}, \text{TM}, \text{SV}\}
\end{equation}
Finally, we concatenate all these to obtain a unified multimodal perceptual feature of the video:
\begin{equation}
\label{eqn:13}
\mathbf{f}_{\text{multimodal}} = [\mathbf{f}_{\text{SE}} \, \parallel \mathbf{f}_{\text{TM}}\, \parallel \mathbf{f}_{\text{SV}} \,].
\end{equation}

\noindent
\textbf{Quality Regressor:}
We employed an MLP regression head, composed of three fully connected layers, to fuse multimodal features for accurate prediction of video quality scores. To mitigate overfitting, we incorporated batch normalization, the GELU activation function, and dropout regularization. For optimization, we used stochastic gradient descent (SGD) with cosine annealing learning rate decay, combined with stochastic weight averaging (SWA), to improve both convergence and generalization. We adopted a composite loss function~\cite{wen2021strong} to enhance accuracy and ranking consistency simultaneously. Details on the loss function settings are provided in the \textit{Supplementary Material}.

\section{Experimental Evaluation}
\label{sec:exp}
% In this section, we first introduce our datasets and evaluation criteria. Next, we provide implementation details and discuss the results. Lastly, we perform ablation studies to analyze the effectiveness of our method.  
\subsection{Datasets And Evaluation Criteria}
\textbf{UGC Datasets:} We validated our proposed method on six mainstream NR-VQA datasets: CVD2014~\cite{nuutinen2016cvd2014}, KoNViD-1k~\cite{hosu2017konstanz}, LIVE-VQC~\cite{sinno2018large}, YouTube-UGC~\cite{wang2019youtube}, LSVQ~\cite{ying2021patch}, and FineVD~\cite{duan2025finevq}. These datasets cover diverse content, authentic distortions, and varying resolutions, broadly representing real-world UGC scenarios. CVD2014 contains authentic videos recorded using various consumer devices. KoNViD-1k comprises 1,200 videos from YFCC100M with moderate resolution and diverse content. LIVE-VQC includes 585 videos with authentic distortions, captured by 80 users across 101 different devices. YouTube-UGC consists of 1,380 videos with varying resolutions, reflecting network-induced quality fluctuations. LSVQ, the largest to date with approximately 39,000 videos, supports data-driven training and generalization evaluation. FineVD is the first large-scale UGC dataset with 800K+ subjective quality ratings and fine-grained descriptions annotated across different quality dimensions. Beyond the general UGC datasets, we also evaluated our method on two specific use cases: gaming videos and short-form videos, to demonstrate its robustness across diverse UGC scenarios. LIVE-YT-Gaming~\cite{yu2023subjective} contains 600 real UGC gaming videos with 18,600 human ratings, capturing the characteristics of synthetic gaming content. KVQ~\cite{lu2024kvq} focuses exclusively on short-form videos with complex editing and realistic processing workflows. We adopted the official partitioning of LSVQ and used an 80\%/20\% train/test split for the remainder, following prior work~\cite{tu2021rapique, li2021unified}.

\noindent
\textbf{Evaluation Metrics:} 
To comprehensively evaluate the performance of the NR-VQA models, we employed two standard statistical metrics: the Spearman Rank Correlation Coefficient (SRCC) and the Pearson Linear Correlation Coefficient (PLCC), which respectively measure the monotonicity and accuracy of the predictions. Both metrics range from 0 to 1, with higher values indicating more accurate rankings and a better numerical fit. To reduce experimental randomness, each set of experiments was repeated 21 times, with the median value taken as the final result~\cite{tu2021rapique}.
% The PLCC is computed after applying a four-parameter nonlinear logistic regression~\cite{seshadrinathan2010study}, which serves to linearise the alignment between the objective predictions and the subjective scores:
\begin{table}[t]
    \centering
    \footnotesize
    \setlength{\tabcolsep}{4pt}
    \caption{Cross-dataset evaluation results, the models are trained on LSVQ and tested on other datasets. The \textcolor{red}{\textbf{red}} entries indicate the best performance.}
    \vspace{-1em}
    \label{tab:cross_lsvq}
    \begin{tabular}{lcccccc}
    \toprule
    \textbf{Test on:} & \multicolumn{2}{c}{\textbf{KoNViD-1k}} & \multicolumn{2}{c}{\textbf{LIVE-VQC}} & \multicolumn{2}{c}{\textbf{YouTube-UGC}}\\
    \cmidrule(lr){2-3} \cmidrule(lr){4-5} \cmidrule(lr){6-7}
    \textbf{Train on: LSVQ} & SRCC & PLCC & SRCC & PLCC & SRCC & PLCC\\
    \midrule
    PVQ\cite{ying2021patch} &0.791 &0.795 &0.770 &0.807 &0.742 &0.754 \\
    FastVQA\cite{wu2023neighbourhood} &0.859 &0.855 &0.823 &0.844 &0.730 &0.747 \\
    DOVER\cite{wu2023exploring} &0.884 &0.883 &0.832 &0.855 &0.777 &0.792\\
    LMM-VQA\cite{ge2024lmm} &0.875 &0.876 &0.831 &0.863 &0.858 &0.877\\
    \rowcolor{gray!15}
    \textbf{CAMP-VQA} &\textcolor{red}{\textbf{0.926}} &\textcolor{red}{\textbf{0.932}} &\textcolor{red}{\textbf{0.919}} &\textcolor{red}{\textbf{0.937}} &\textcolor{red}{\textbf{0.880}} &\textcolor{red}{\textbf{0.898}}\\
    \bottomrule
    \end{tabular}
\vspace{-1em}
\end{table}
\begin{table}[t]
    \centering
    \footnotesize
    \setlength{\tabcolsep}{1.5pt}
    \caption{Performance comparison on UGC video datasets for gaming and short-form use cases. The \textcolor{red}{\textbf{red}} entries indicate the best performance.}
    \vspace{-1em}
    \label{tab:ugc_other}
    \begin{tabular}{@{}llcccccccccccc@{}}
    \toprule 
    \multicolumn{2}{l}{\textbf{Target Quality Dataset}} & \multicolumn{2}{c}{\textbf{LIVE-YT-Gaming}} & \multicolumn{2}{c}{\textbf{KVQ}} \\
    \cmidrule(lr){1-2} \cmidrule(lr){3-4} \cmidrule(lr){5-6}
    Model & \shortstack{Extra\\Data} & SRCC & PLCC & SRCC & PLCC \\
    \midrule
    FastVQA\cite{wu2023neighbourhood} &LSVQ &0.869 &0.880 &0.832 &0.834\\ 
    DOVER\cite{wu2023exploring} &LSVQ &0.852 &0.868 &0.833 &0.837\\
    KSVQE\cite{lu2024kvq} &KVQ &N/A &N/A &0.867 &0.869\\
    LMM-VQA\cite{ge2024lmm} &LSVQ &0.816 &0.801 &N/A &N/A\\
    FineVQ\cite{duan2025finevq} &FineVD &\textcolor{red}{\textbf{0.912}} &0.926 &N/A &N/A\\
    \rowcolor{gray!15}
    \textbf{CAMP-VQA} &None &0.903 &0.922 &0.956 &0.958\\
    \rowcolor{gray!15}
    \textbf{CAMP-VQA (\textit{cross-dataset})} &LSVQ &0.864 &0.884 &0.811 &0.810\\
    \rowcolor{gray!15}
    \textbf{CAMP-VQA (\textit{w/ fine-tune})} &LSVQ &0.905 &\textcolor{red}{\textbf{0.942}} &\textcolor{red}{\textbf{0.967}} &\textcolor{red}{\textbf{0.967}}\\
    \bottomrule
    \end{tabular}
\vspace{-2em}
\end{table}

\subsection{Implementation Details}
Our method was implemented using Python 3.10 and the PyTorch framework, and all experiments were carried out using one NVIDIA RTX 6000 Ada GPU. For feature extraction in SEE, after generating quality-aware captions using BLIP-2 with a Flan-T5-XL decoder, the captions and sampled video frames were subsequently encoded using CLIP (ViT-B/32) to derive semantic embeddings. For TME, we used SlowFast pre-trained on Kinetics-400~\cite{kay2017kinetics} to extract temporal features. For SVE, spatial features were extracted using Swin-Large pre-trained on ImageNet-22k~\cite{krizhevsky2012imagenet}. The quality regressor is a three-layer MLP with hidden layers of 256 and 128 dimensions, and an output layer producing a scalar score. A dropout rate of 0.1 is applied between layers to prevent overfitting. We trained the quality regressor using the SGD optimizer with a batch size of 256 for 50 epochs, an initial learning rate of 1e-1, and a weight decay of 0.005. To enhance generalization, 10-fold cross-validation was applied on the large-scale LSVQ dataset. For training/fine-tuning on smaller-scale datasets, we trained for 200 epochs with a learning rate of 1e-2 and a weight decay of 0.0005, without cross-validation. In later training stages, SWA and early stopping with a patience of 5 were applied. The optimal model parameters were selected based on the minimum root mean squared error between the predicted scores and the ground truth on the validation set.

\begin{table*}[h]
    \centering
    \footnotesize
    \setlength{\tabcolsep}{2pt}
    \caption{Performance of the evaluated NR-VQA models and the proposed CAMP-VQA on FineVD is reported in terms of quality scoring across different quality dimensions. $\bar{\mathbf{e}}^{\text{img}}$, $\bar{\mathbf{e}}^{\text{qlt}}$, and $\bar{\mathbf{e}}^{\text{art}}$ denote the use of only image, quality, and artifact embeddings of video clips, respectively. The \textcolor{red}{\textbf{red}} and \textcolor{blue}{\textbf{blue}} entries indicate the 1st and 2nd best on each quality dimension, respectively.}
    \vspace{-1em}
    \label{tab:ablation_1}
    \begin{tabular}{ccccccccccccccccccc}
    \toprule
    \multicolumn{1}{r}{\textbf{Dimension:}} &  \multicolumn{3}{c}{\textbf{Color}} & \multicolumn{3}{c}{\textbf{Noise}} & \multicolumn{3}{c}{\textbf{Artifact}} & \multicolumn{3}{c}{\textbf{Blur}}  & \multicolumn{3}{c}{\textbf{Temporal}} & \multicolumn{3}{c}{\textbf{Overall}}\\
    \cmidrule(lr){2-4}\cmidrule(lr){5-7}\cmidrule(lr){8-10}\cmidrule(lr){11-13}\cmidrule(lr){14-16}\cmidrule(lr){17-19}
    \textbf{Metrics} & SRCC & KRCC & PLCC & SRCC & KRCC & PLCC & SRCC & KRCC & PLCC & SRCC & KRCC & PLCC & SRCC & KRCC & PLCC & SRCC & KRCC & PLCC\\
    \midrule
    \textbf{VIDEVAL} &0.692 & 0.503 & 0.694 & 0.691 & 0.503 & 0.651 & 0.744 & 0.550 & 0.737 & 0.761 & 0.565 & 0.764 & 0.717 & 0.531 & 0.712 & 0.731 & 0.537 & 0.731\\
    \textbf{VSFA} &0.762 & 0.567 & 0.784 & 0.764 & 0.571 & 0.728 & 0.801 & 0.608 & 0.817 & 0.777 & 0.586 & 0.800 & 0.728 & 0.536 & 0.702 & 0.773 & 0.583 & 0.793\\
    \textbf{FAST-VQA} &0.802 & 0.608 & 0.818 & 0.809 & 0.621 & 0.776 & 0.818 & 0.630 & 0.833 & 0.835 & 0.650 & 0.851 & 0.756 & 0.561 & 0.739 & 0.835 & 0.648 & 0.847\\
    \textbf{DOVER} &0.824 & 0.631 & 0.831 & 0.802 & 0.606 & 0.742 & 0.827 & 0.634 & 0.829 & 0.840 & 0.650 & 0.836 & 0.766 & 0.570 & 0.757 & 0.842 & 0.652 & 0.839\\
    \textbf{FineVQ} &\textcolor{blue}{\textbf{0.850}} &\textcolor{blue}{\textbf{0.667}} &\textcolor{blue}{\textbf{0.853}} &\textcolor{blue}{\textbf{0.844}} &\textcolor{blue}{\textbf{0.661}} &\textcolor{blue}{\textbf{0.799}} &\textcolor{blue}{\textbf{0.885}} &\textcolor{blue}{\textbf{0.711}} &\textcolor{blue}{\textbf{0.892}} &\textcolor{blue}{\textbf{0.871}} &\textcolor{blue}{\textbf{0.696}} &\textcolor{blue}{\textbf{0.883}} &\textcolor{blue}{\textbf{0.809}} &\textcolor{blue}{\textbf{0.617}} &\textcolor{blue}{\textbf{0.760}} &\textcolor{blue}{\textbf{0.883}} &\textcolor{blue}{\textbf{0.712}} &\textcolor{blue}{\textbf{0.889}}\\
    \hdashline
    \noalign{\vskip 2pt}
    \rowcolor{gray!15}
    \textbf{$\bar{\mathbf{e}}^{\text{img}}$} &0.804 &0.612 &0.828 &0.763 &0.569 &0.727 &0.806 &0.612 &0.814 &0.802 &0.607 &0.814 &0.723 &0.532 &0.721 &0.804 &0.611 &0.817\\
    \rowcolor{gray!15}
    \textbf{$\bar{\mathbf{e}}^{\text{qlt}}$} &0.768 &0.579 &0.804 &0.745 &0.556 &0.767 &0.761 &0.559 &0.827 &0.768 &0.565 &0.828 &0.723 &0.535 &0.736 &0.816 &0.616 &0.869\\
    \rowcolor{gray!15}
    \textbf{$\bar{\mathbf{e}}^{\text{art}}$} &0.750 &0.541 &0.765 &0.764 &0.556 &0.736 &0.787 &0.577 &0.807 &0.778 &0.563 &0.809 &0.736 &0.527 &0.713 &0.812 &0.594 &0.840\\
    \rowcolor{gray!15}
    \textbf{CAMP-VQA} &\textcolor{red}{\textbf{0.893}} &\textcolor{red}{\textbf{0.712}} &\textcolor{red}{\textbf{0.884}} &\textcolor{red}{\textbf{0.873}} &\textcolor{red}{\textbf{0.680}} &\textcolor{red}{\textbf{0.819}} &\textcolor{red}{\textbf{0.902}} &\textcolor{red}{\textbf{0.725}} &\textcolor{red}{\textbf{0.904}} &\textcolor{red}{\textbf{0.906}} &\textcolor{red}{\textbf{0.730}} &\textcolor{red}{\textbf{0.907}} &\textcolor{red}{\textbf{0.851}} &\textcolor{red}{\textbf{0.658}} &\textcolor{red}{\textbf{0.831}} &\textcolor{red}{\textbf{0.919}} &\textcolor{red}{\textbf{0.749}} &\textcolor{red}{\textbf{0.923}}\\
    \bottomrule
    \end{tabular}
\vspace{-2em}
\end{table*}

\begin{table}[t]
    \centering
    \footnotesize
    \setlength{\tabcolsep}{4.5pt}
    \caption{Ablation study on the effect of different component semantic embeddings on KoNViD-1k and FineVD: image, quality, artifact, and content ($\bar{\mathbf{e}}^{\text{img}}$, $\bar{\mathbf{e}}^{\text{qlt}}$, $\bar{\mathbf{e}}^{\text{art}}$, $\mathrm{content}_{\text{embs}}$). The \textcolor{red}{\textbf{red}} entries indicate the best performance.}
    \vspace{-1em}
    \label{tab:ablation_2}
    \begin{tabular}{cccccccc}
    \toprule
    \multicolumn{4}{r}{\textbf{Test Set:}} & \multicolumn{2}{c}{\textbf{KoNViD-1k}} & \multicolumn{2}{c}{\textbf{FineVD}}\\
    \cmidrule(lr){5-6} \cmidrule(lr){7-8}
    $\bar{\mathbf{e}}^{\text{img}}$ & $\bar{\mathbf{e}}^{\text{qlt}}$ & $\bar{\mathbf{e}}^{\text{art}}$ & $\mathrm{content}_{\text{embs}}$ & SRCC & PLCC & SRCC & PLCC\\
    \midrule
    % $\mathrm{img}_{\text{embs}}$ & $\mathrm{quality}_{\text{embs}}$ & $\mathrm{artifact}_{\text{embs}}$
    \checkmark &            &            &             & 0.778 & 0.804 &0.804 &0.817\\
               & \checkmark &            &             & 0.631 & 0.792 &0.816 &0.869\\
               &            & \checkmark &             & 0.735 & 0.763 &0.812 &0.840\\
               &            &            & \checkmark  & 0.409 & 0.451 &0.401 &0.409\\
    \checkmark & \checkmark &            &             & 0.830 & 0.871 &0.899 &0.911\\
    \rowcolor{gray!15}
    \checkmark & \checkmark & \checkmark &             & \textcolor{red}{\textbf{0.903}} & \textcolor{red}{\textbf{0.922}} &\textcolor{red}{\textbf{0.901}} &\textcolor{red}{\textbf{0.919}}\\
    \checkmark & \checkmark & \checkmark & \checkmark  & 0.892 & 0.919 &0.896 &0.917\\
    \bottomrule
    \end{tabular}
\vspace{-2em}
\end{table}

\subsection{Performance Comparison}
We evaluated the proposed CAMP-VQA model, on six mainstream UGC benchmark datasets. The experimental testing comprised:
% : CVD2014~\cite{nuutinen2016cvd2014}, KoNViD-1k~\cite{hosu2017konstanz}, LIVE-VQC~\cite{sinno2018large}, YouTube-UGC~\cite{wang2019youtube}, and LSVQ~\cite{ying2021patch}. 
\begin{enumerate}[leftmargin=*, itemsep=2pt, topsep=2pt, parsep=0pt]
    \item Training and testing on each target dataset, referred to as \textit{intra-dataset} experiments.
    \item Pre-training the model on LSVQ, followed by fine-tuning on the target datasets (denoted as \textit{w/ fine-tune}), aimed to assess the model's transferability and adaptability.
    % adaptation capabilities.
\end{enumerate}

\begin{table*}[h]
    \centering
    \footnotesize
    \setlength{\tabcolsep}{5pt}
    \caption{Ablation study on semantic, temporal, and spatial feature ($\mathrm{\mathbf{f}_{\text{SE}}}$,  $\mathrm{\mathbf{f}_{\text{TM}}}$, $\mathrm{\mathbf{f}_{\text{SV}}}$) extractors. \textcolor{red}{\textbf{Red}} entries denote best performance.}
    \vspace{-1em}
    \label{tab:ablation_3}
    \begin{tabular}{ccccccccccccccccc}
    \toprule
    \multicolumn{3}{r}{\textbf{Test Set:}} &  \multicolumn{2}{c}{\textbf{CVD2014}} & \multicolumn{2}{c}{\textbf{KoNViD-1k}} & \multicolumn{2}{c}{\textbf{LIVE-VQC}} & \multicolumn{2}{c}{\textbf{YouTube-UGC}}  & \multicolumn{2}{c}{\textbf{LSVQ\(_{\text{test}}\)}} & \multicolumn{2}{c}{\textbf{LSVQ\(_{\text{1080p}}\)}} & \multicolumn{2}{c}{\textbf{FineVD}}\\
    \cmidrule(lr){4-5}\cmidrule(lr){6-7}\cmidrule(lr){8-9}\cmidrule(lr){10-11}\cmidrule(lr){12-13}\cmidrule(lr){14-15}\cmidrule(lr){16-17}
    $\mathrm{\mathbf{f}_{\text{SE}}} $ & $\mathrm{\mathbf{f}_{\text{TM}}}$ & $\mathrm{\mathbf{f}_{\text{SV}}}$ & SRCC & PLCC & SRCC & PLCC & SRCC & PLCC & SRCC & PLCC & SRCC & PLCC & SRCC & PLCC & SRCC & PLCC\\
    % $\mathrm{semantic} $ & $\mathrm{temporal}$ & $\mathrm{spatial}$
    \midrule
    \checkmark &            &             &0.916 &0.924 &0.902 &0.922 &0.905 &0.935 &0.887 &0.919 &0.898 &0.921 &0.882 &0.903 &0.902 &0.919\\
               & \checkmark &             &0.825 &0.835 &0.770 &0.782 &0.739 &0.758 &0.710 &0.718 &0.824 &0.827 &0.728 &0.762 &0.719 &0.725\\
               &            & \checkmark  &0.858 &0.887 &0.861 &0.860 &0.777 &0.820 &0.817 &0.833 &0.867 &0.868 &0.754 &0.804 &0.838 &0.840\\
    \checkmark & \checkmark &             &0.923 &0.941 &0.909 &0.921 &0.917 &0.936 &0.885 &0.910 &0.910 &0.928 &0.901 &0.912 &0.907 &0.918\\
    \rowcolor{gray!15}
    \checkmark & \checkmark & \checkmark  &\textcolor{red}{\textbf{0.933}} &\textcolor{red}{\textbf{0.944}} &\textcolor{red}{\textbf{0.927}} &\textcolor{red}{\textbf{0.936}} &\textcolor{red}{\textbf{0.922}} &\textcolor{red}{\textbf{0.940}} &\textcolor{red}{\textbf{0.901}} &\textcolor{red}{\textbf{0.920}} &\textcolor{red}{\textbf{0.920}} &\textcolor{red}{\textbf{0.933}} &\textcolor{red}{\textbf{0.908}} &\textcolor{red}{\textbf{0.920}} &\textcolor{red}{\textbf{0.919}} &\textcolor{red}{\textbf{0.923}}\\
    \bottomrule
    \end{tabular}
\vspace{-2em}
\end{table*}

As shown in Table~\ref{tab: ComparisonToSoA}, we compared our method against a range of SOTA models, including traditional hand-crafted models (e.g., VIDEVAL), frame-based deep learning models (e.g., VSFA), fragment-based deep learning models (e.g., DOVER), and recently proposed large multimodal models (e.g., LMM-VQA). For the \textit{intra-dataset} experiments, CAMP-VQA demonstrated outstanding performance across all datasets. Compared to the traditional method VIDEVAL, it achieved average improvements of $\Delta \mathrm{SRCC}=0.177$ and $\Delta \mathrm{PLCC}=0.183$ across datasets. Compared to the frame-based method VSFA, the average gains were $\Delta \mathrm{SRCC}=0.149$ and $\Delta \mathrm{PLCC}=0.149$. On both the LSVQ$_{\text{test}}$ and LSVQ$_{\text{1080p}}$ test sets, CAMP-VQA showed the best performance, with especially notable gains on the high-resolution 1080p subset, outperforming the second-best LMM-VQA, by $\Delta \mathrm{SRCC}=0.017$ and $\Delta \mathrm{PLCC}=0.021$ ($^*$ The \textit{w/ fine-tune} results are identical to the inter-dataset results since both are trained and tested on LSVQ).

In the \textit{w/ fine-tune} scenario, CAMP-VQA achieved SOTA performance across all target datasets. Compared to the fragment-based method DOVER, the improvements were $\Delta \mathrm{SRCC}=0.065$ and $\Delta \mathrm{PLCC}=0.065$. Against the current SOTA multimodal method LMM-VQA, CAMP-VQA achieved performance gains of $\Delta \mathrm{SRCC}=0.022$ and $\Delta \mathrm{PLCC}=0.028$. For instance, on KoNViD-1k and LIVE-VQC, it attained PLCC of 0.944 and 0.946, respectively, significantly surpassing previous SOTA NR-VQA models. On the multi-resolution, real-user datasets YouTube-UGC and FineVD, CAMP-VQA also led with a PLCC of 0.928 and 0.933, demonstrating strong perceptual capability across a wide range of quality conditions. In summary, in both experimental setups, CAMP-VQA consistently achieved SOTA performance without the need for fine-grained manual annotations. These results validate the effectiveness of the proposed quality-aware caption generation and multimodal video feature fusion, underscoring its strong potential for deployment.

% main table with all the other vqa models (CVD2014, Kovid-1k, live-vqa, youtube-ugc, lsvq, finevd)
% \textit{intra-dataset} experiment:
% For VIDEVAL: SRCC = 0.766 + 0.807 + 0.773 + 0.781 + 0.794 + 0.545 + 0.731 = 5.197, PLCC = 0.806 + 0.792 + 0.775 + 0.793 + 0.783 + 0.554 + 0.731 = 5.234
               % SRCC = 0.742, PLCC = 0.748, ∆SRCC = 0.919 − 0.742 = +0.177，∆PLCC = 0.931 − 0.748 = +0.183
% For VSFA: SRCC = 0.870 + 0.773 + 0.773 + 0.724 + 0.801 + 0.675 + 0.773 = 5.389, PLCC = 0.868 + 0.775 + 0.795 + 0.743 + 0.796 + 0.704 + 0.793 = 5.474
            % SRCC = 0.770, PLCC = 0.782, ∆SRCC = 0.919 − 0.770 = +0.149，∆PLCC = 0.931 − 0.782 = +0.149
% For LMM-VQA with LSVQ\_1080p: ∆SRCC = 0.908-0.891 = 0.017, ∆PLCC = 0.920-0.899 = 0.021
% \textit{w/ fine-tune} experiment:
% For DOVER: SRCC = 0.858 + 0.909 + 0.860 + 0.890 + 0.888 + 0.795 + 0.842 = 6.042, PLCC = 0.881 + 0.906 + 0.875 + 0.891 + 0.889 + 0.830 + 0.839 = 6.111
             % SRCC = 0.863, PLCC = 0.873, ∆SRCC = 0.928 − 0.863 = +0.065，∆PLCC = 0.938 − 0.873 = +0.065
% For LMM-VQA: SRCC = 0.929 + 0.891 + 0.901 + 0.916 + 0.891 = 4.528, PLCC = 0.930 + 0.903 + 0.897 + 0.919 + 0.899 = 4.548
               % SRCC = 0.906, PLCC = 0.910, ∆SRCC = 0.928 − 0.906 = +0.022，∆PLCC = 0.938 − 0.910 = +0.028
% CAMP-VQA:               
% For CAMP-VQA: SRCC = 0.933 + 0.927 + 0.922 + 0.901 + 0.920 + 0.908 + 0.919 = 6.43, PLCC = 0.944 + 0.936 + 0.940 + 0.920 + 0.933 + 0.920 + 0.923 = 6.516
                % SRCC = 0.919, PLCC = 0.931
% For CAMP-VQA (fine-tune): SRCC = 0.966 + 0.930 + 0.934 + 0.912 + 0.920 + 0.908 + 0.924 = 6.494, PLCC = 0.964 + 0.944 + 0.946 + 0.928 + 0.933 + 0.920 + 0.933 = 6.568
                % SRCC = 0.928, PLCC = 0.938
                
\subsection{Cross-dataset Evaluation}
Table~\ref{tab:cross_lsvq} presents cross-dataset evaluation of CAMP-VQA after training on LSVQ, showing SOTA performance on KoNViD-1k, LIVE-VQC, and YouTube-UGC. Specifically, our model achieved a PLCC of 0.932 on KoNViD-1k, marking an improvement of approximately 9\% over FAST-VQA. On LIVE-VQC, the SRCC is 0.919, outperforming LMM-VQA by approximately 11\%. It also surpassed all competing methods on YouTube-UGC, further demonstrating the strong cross-dataset generalization of our model. In the specific UGC use case, CAMP-VQA (\textit{w/ fine-tune}) also achieved leading results on the LIVE-YT-Gaming dataset, which focuses on UGC gaming content, and the KVQ dataset, which targets short video scenarios. As shown in Table~\ref{tab:ugc_other}, our model reached a PLCC of 0.942 on LIVE-YT-Gaming, outperforming FineVQ by approximately 1.73\%. On KVQ, it achieved both SRCC and PLCC scores of 0.967, representing improvements of approximately 11.5\% and 11.3\% over KSVQE, respectively. 
% Notably, the model also showed strong performance under the intra-dataset setting.

% lsvq Cross-dataset Evaluation:
% For FAST-VQA with KoNViD-1k: PLCC% = (0.932-0.855)/0.855 = 0.09006 ≈ 9%
% For LMM-VQA with LIVE-VQC: PLCC% = (0.937-0.863)/0.863 = 0.0857 ≈ 8.6%  SRCC% = (0.919-0.831)/0.831 = 0.1059 ≈ 11%
% For FineVQ with LIVE-YT-Gaming: % = (0.942-0.926)/0.926 = 0.01728 ≈ +1.73%
% For KSVQE with KVQ: SRCC% = (0.967-0.867)/0.867 = 0.1153 ≈ 11.5%, PLCC% = (0.967-0.869)/0.869 = 0.1128 ≈ 11.3%

\subsection{Ablation Studies}
We explored the impact of each key component on performance: semantic embeddings, prompt hints, and multimodal features.

\noindent
\textbf{On Semantic Embeddings:} 
Building on our proposed quality-aware captioning, we explored the impact of generated captions on video quality scoring through the semantic embeddings of sampled frames. Specifically, we assessed image ($\bar{\mathbf{e}}^{\text{img}}$), quality ($\bar{\mathbf{e}}^{\text{qlt}}$), and artifact ($\bar{\mathbf{e}}^{\text{art}}$) embeddings, both independently and in combination. Table~\ref{tab:ablation_1} summarizes the performance of three semantic embeddings and CAMP-VQA, compared to other NR-VQA models on FineVD. The results are reported across five quality dimensions: color, noise, artifact, blur, and temporal, as well as overall MOS. These embeddings can serve as substitutes of fine-grained annotations and exhibit correlations aligned with perceptual quality dimensions. Image embeddings capture frame-level semantics, correlating well in color (SRCC=0.804, PLCC=0.828). As text inputs, quality embeddings describe general perceptual degradations and perform best on overall MOS among single embeddings (SRCC=0.816, PLCC=0.869). Artifact embeddings represent captioned compression distortions, capture artifact-related cues, and perform better on artifact and blur (SRCC=0.807, 0.809) than on other dimensions. All semantic embeddings show lower prediction accuracy on the temporal dimension. We address this by integrating spatio-temporal features into CAMP-VQA, which consistently achieves the best performance across all dimensions, with an average SRCC of 0.891, KRCC of 0.709, and PLCC of 0.878. We provide example videos along with the predicted scores of CAMP-VQA and ground-truth MOS of FineVD across different quality dimensions in \textit{Supplementary Material}.
% Semantic embeddings capture information aligned with fine-grained labels, enabling effective supervision of degradation dimensions without manual annotation.

As depicted in Table~\ref{tab:ablation_2}, we employed SEE to extract semantic embeddings of different components and evaluated their effectiveness on KoNViD-1k and FineVD. In addition, we analyzed the impact of content embeddings (descriptions of video content). The results indicate that the model achieves the best performance when the three quality-aware dimensions (image, quality, and artifact) are combined, yielding SRCC and PLCC scores of 0.903 and 0.922 on KoNViD-1k, and 0.901 and 0.919 on FineVD. In contrast, using only image embeddings results in lower SRCC scores of 0.778 and 0.804, while using only quality embeddings yields 0.631 and 0.816, respectively. These observations suggest that a single quality-aware dimension is inadequate for capturing spatio-temporal quality features. In comparison, fusing multiple semantic embeddings enables more effective modeling of the relationship between video distortions and perceived quality, thereby improving the accuracy of the VQA model. Notably, content embeddings were found to degrade quality prediction. When used alone, the SRCC dropped to 0.409 and 0.401. Adding content information to the best-performing configuration also led to a performance decline.
% ∆SRCC = (0.893 + 0.873 + 0.902 + 0.906 + 0.851 + 0.919)/6 = 0.891
% ∆KRCC = (0.712 + 0.680 + 0.725 + 0.730 + 0.658 + 0.749)/6 = 0.709
% ∆PLCC = (0.884 + 0.819 + 0.904 + 0.907 + 0.831 + 0.923)/6 = 0.878
% This suggests that for VQA tasks primarily focused on complex distortions and artifacts, the correlation between video content and low-level visual quality is relatively weak.

\noindent
\textbf{On Quality-related Prompt Hints:} 
Similarly to the above, we extracted quality-aware semantic embeddings, combined the three dimensions, and tested on KoNViD-1k. Upon incorporating quality-related hints, SRCC and PLCC improved from 0.767 to 0.900 and from 0.790 to 0.921, respectively. These results demonstrate that the proposed quality-related prompts, derived from video metadata, effectively guide BLIP-2 in generating captions that align more closely with human perception. Consequently, the model’s sensitivity to complex distortions is markedly enhanced, leading to improved prediction performance.
% Table~\ref {tab: ablation} provides an analysis of the importance of quality-related prompt hints. 
% It further reinforces the effectiveness and applicability of prompt engineering in multimodal NR-VQA models.

\noindent
\textbf{On Multimodal Feature Extractors:} 
Table~\ref{tab:ablation_3} demonstrates the impact of different modalities on performance across six UGC datasets. The proposed multimodal feature fusion consistently yields optimal performance. For example, on LIVE-VQC, integrating all three modalities resulted in improvements of approximately 24.8\% in SRCC and 24\% in PLCC compared to using only temporal features. Similar substantial enhancements were observed across the remaining datasets. Semantic features contribute to high-level perception, temporal features capture dynamic distortion changes within the video, and spatial features reflect image texture structures and local distortions. These three features complement one another, forming a more robust and discriminative representation of video quality. Including only a single modality leads to performance decline, further highlighting the vital role of multimodal video feature fusion.
% LIVE-VQC: SRCC% = (0.922-0.739)/0.739 = 0.2476 ≈ 24.8%, PLCC% = (0.940-0.758)/0.758 = 0.2401055409 ≈ 24%  

\section{Conclusion}
\label{sec:con}
In this paper, we propose a novel NR-VQA model that uses a VLM to obtain artifact semantics annotations and combines those with spatio-temporal features. Particularly, our method introduces quality-aware caption generation, enabling automatic extraction of fine-grained, quality-related captions as substitutes for manual annotations of visual artifacts. Additionally, we establish a unified multimodal VQA framework that integrates SVE, TME, and SEE feature extractors, resulting in the extraction of quality-aware video representations through feature fusion. Our model achieves SOTA performance across all UGC datasets and consistently demonstrates robust cross-dataset generalization, attaining an average SRCC of 0.928 and PLCC of 0.938. Future research will focus on incorporating high-level semantic understanding, aiming to unify video question answering and video quality assessment.

\small
\bibliographystyle{ieeenat_fullname}
\bibliography{ref}

% Appendix
\newpage

\appendix
\section{Supplementary Material}
This document is the supplementary material for our model CAMP-VQA. We provide further explanation and details on frame difference fragmentation, quality-aware caption generation, spatio-temporal feature extraction, and loss functions. Additional figures are included to illustrate the quality-related prompt settings and to demonstrate the effectiveness of the proposed CAMP-VQA in predicting scores across different quality dimensions.

\section{Frame Difference Fragmentation}
\begin{figure}[htbp]
    \centering
    \includegraphics[width=1\linewidth]{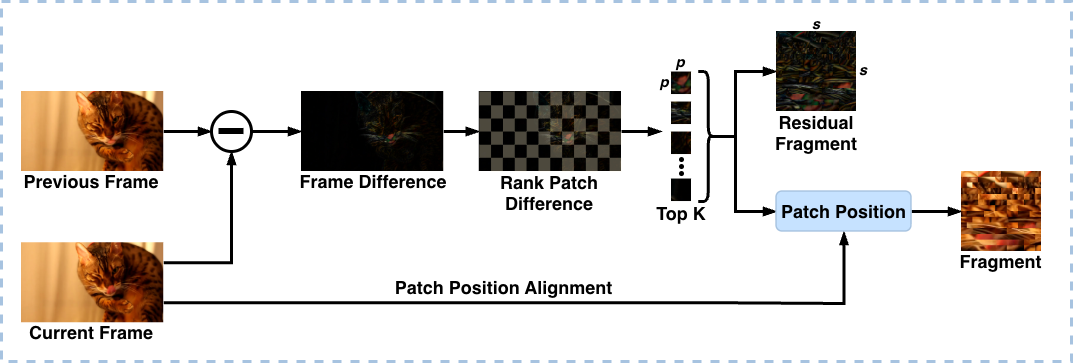}
    \caption{Frame difference fragmentation (FDF) module.}
    \label{fig: fragmentation}
\end{figure}

Regarding patch selection, we compute the inter-frame absolute differences for all patches and rank them accordingly. The fragmentation details are shown in (Fig.~\ref{fig: fragmentation}).
The number of selected patches $K$ is calculated based on the target model input size $s \times s$ and each patch size $p \times p$:
\begin{equation}
\label{eqn:a01}
K = \tfrac{s^2}{p^2}.
\end{equation}
We then select the top $K$ patches from the ranked list based on the sum of absolute inter-frame differences. To ensure spatial consistency between the extracted frame and residual fragments, each selected patch in the residual is paired with a corresponding fragment extracted from the same pixel position in $F_t$. For a target size of $s = 224$, we use a patch size of $p = 16$, resulting in $K = 196$. This method generalizes across resolutions, as fragmentation operates directly on the original frames. It is important to note that similar residual intensities can arise from different types of motion, which are likely to be associated with distinct distortion types.

\section{Semantic-aligned Feature Extraction}
\begin{figure}[htbp]
    \centering
    \includegraphics[width=1\linewidth]{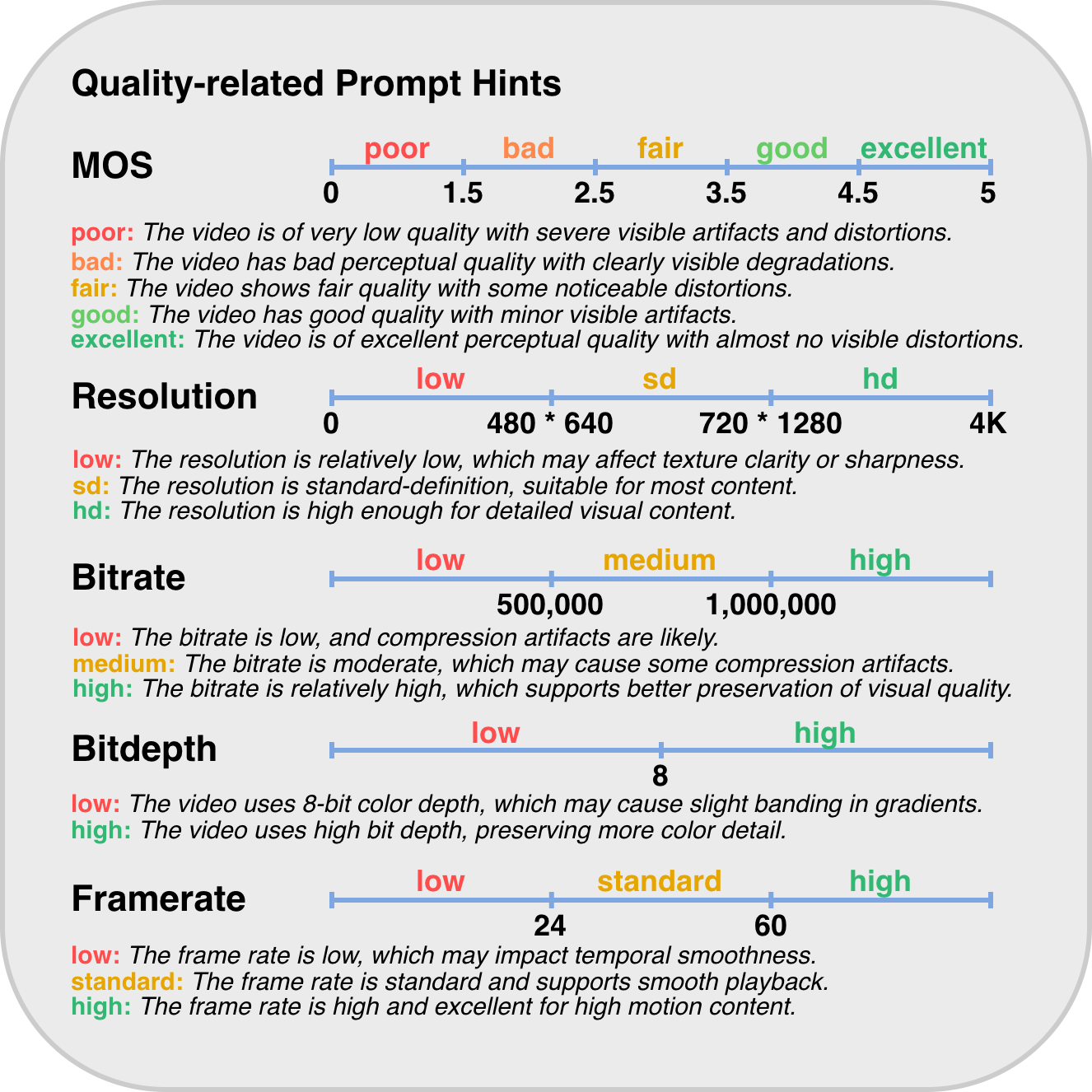}
    \caption{Quality-related prompt hints derived from video metadata.}
    \label{fig: qualhint}
\end{figure}
Based on video metadata across multiple dimensions, we derive a set of quality-related prompt hints to guide the input prompt, as illustrated in Fig.~\ref{fig: qualhint}. The figure shows how metadata such as resolution, bitrate, and frame rate are mapped to qualitative levels based on their values, each accompanied by descriptive hints. These mappings provide a structured way to integrate perceptual video quality factors into prompt construction for specific videos. Notably, we avoid disclosing ground-truth MOS scores. During training, quality-related hints are generated from quantized quality levels (ranging from poor to excellent) derived from metadata, guiding prompt text generation without including explicit scores in the descriptions. During inference, the quality level is set to prediction mode, eliminating the need for MOS.

\begin{figure*}[htbp]
    \centering
    \includegraphics[width=.75\linewidth]{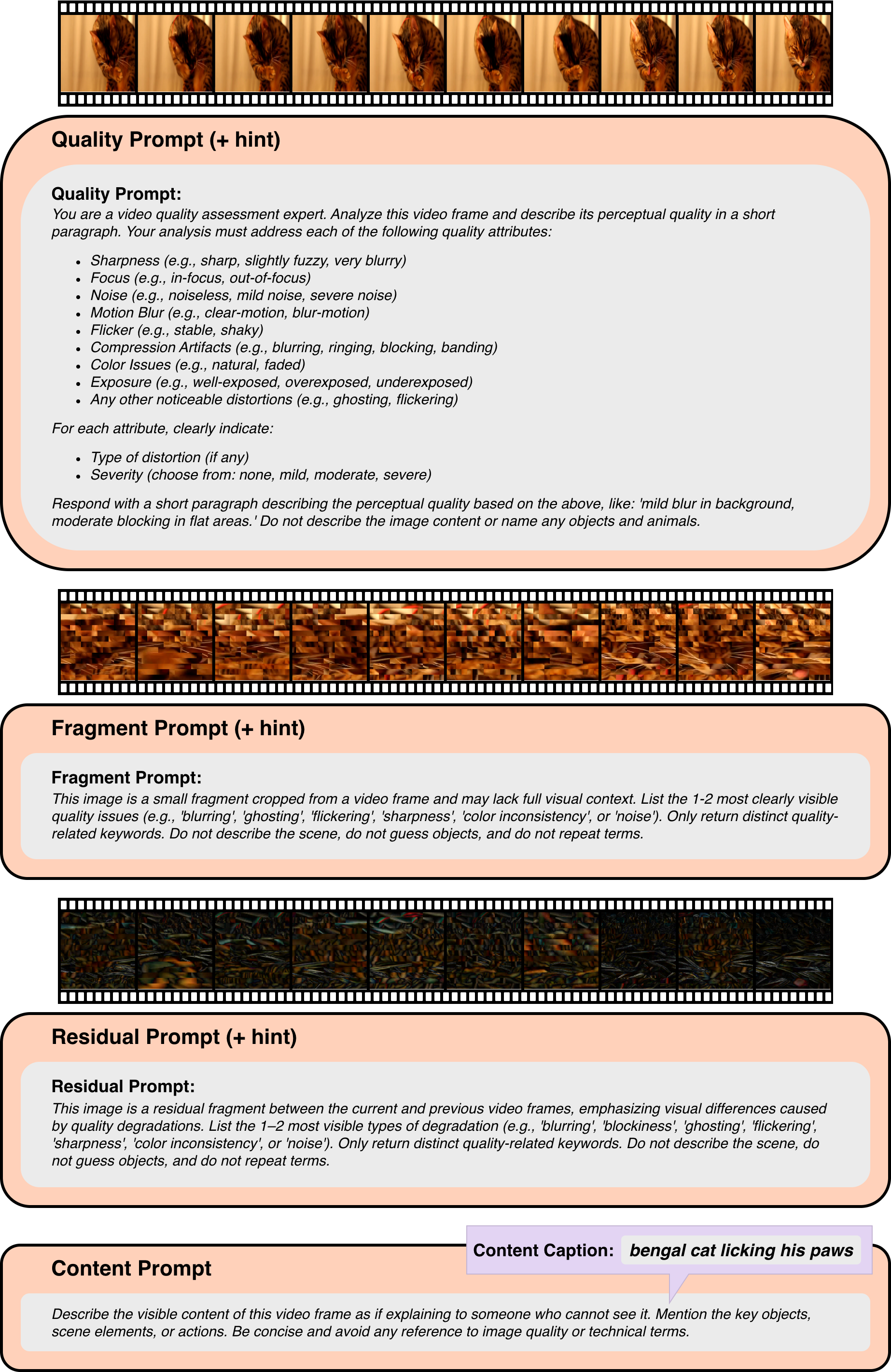}
    \caption{Different quality-aware prompt settings: quality prompt, fragment prompt, and residual prompt. We also include a content prompt for the ablation study.}
    \label{fig: qualprompt}
\end{figure*}

Upon these hints, we design different quality-aware prompt settings, as illustrated in Fig.~\ref{fig: qualprompt}. The quality prompt is used to extract general perceptual captions at the frame level across multiple visual attributes. The fragment prompt focuses on localized degradations, and the residual prompt captures temporal differences between consecutive frames, jointly providing artifact captions at the fragment level. In the ablation study, we also include a content prompt to analyze the role of content descriptions in low-level video quality assessment. Each prompt is tailored to capture a different aspect of perceptual quality.

\section{Spatio-temporal Feature Extraction} 
\subsection{Spatial Vision Extractor} 
The Swin Transformer (SwinT)~\cite{liu2021swin} employs a hierarchical sliding window mechanism and utilizes local self-attention to capture spatial features and long-range dependencies within images. To make full use of its strong spatial modeling capacity, we discard the classification head and retain only the backbone as our Spatial Vision Extractor (SVE) module.

Input frames of a video clip in the form of a tensor $\mathbf{X_\text{frame}} \in \mathbb{R}^{N \times C \times H \times W}$, where $N$ indicates the number of frames, $C$ is the number of channels, and $H$ and $W$ are the spatial dimensions. Each frame tensor is encoded by the SwinT backbone network $\phi_s$, and compact spatial representations are generated through GAP:
\begin{equation}
\label{eqn:a01}
z_{\text{swint}} = \text{GAP}(\phi_\text{swin}(\mathbf{X_\text{frame}})) \in \mathbb{R}^{N \times d_s},
\end{equation}
where $d_s$ denotes the dimension of the spatial features.
SVE module captures the spatial structural information of video frames, enabling spatial awareness for subsequent temporal modeling and multimodal fusion.

\subsection{Temporal Motion Extractor} 
To enhance the model's capacity for spatio-temporal feature modeling in videos, we designed a Temporal Motion Extractor (TME) module based on the SlowFast~\cite{feichtenhofer2019slowfast} architecture. TME uses a dual pathway to extract features across different temporal scales, enabling effective capture of quality fluctuations in videos while maintaining efficiency. 

We construct two pathways with different temporal divisions by using an input frame sequence tensor $\mathbf{X_\text{sequence}} \in \mathbb{R}^{B \times C \times T \times H \times W}$, where $B$ is batch size, $C$ is channels, $T$ is temporal length, and $H, W$ are the height and width. The slow pathway extracts low-frequency changes over longer temporal spans by downsampling at a rate of $r = \frac{1}{4}$, denoted as $X_{\text{slow}} = \text{Sample}(\mathbf{X_\text{sequence}}, r)$. The fast pathway retains the original frame rate, i.e., $X_{\text{fast}} = \mathbf{X_\text{sequence}}$, and serves to capture rapidly high-frequency details. The two pathways are fed into feature extractors $\phi_\text{s}$ and $\phi_\text{f}$, respectively. Spatio-temporal GAP is then applied to produce compact feature representations:
\begin{equation}
\label{eqn:a02}
\mathbf{z}_{\text{slow}} = \mathrm{GAP}(\phi_\text{s}(\mathbf{X}_{\text{slow}})),\ 
\mathbf{z}_{\text{fast}} = \mathrm{GAP}(\phi_\text{f}(\mathbf{X}_{\text{fast}})).
\end{equation}
Finally, the two feature vectors are concatenated to form the temporal feature:
\begin{equation}
\label{eqn:a03}
\mathbf{z_{slowfast}} = \left[ \mathbf{z}_{\text{slow}} ; \mathbf{z}_{\text{fast}} \right] \in \mathbb{R}^{B \times (d_s + d_f)},
\end{equation}
where $d_s$ and $d_f$ are the feature dimensions of the slow and fast channels, respectively. This module extracts temporal motion features at different time scales, enhancing the model's sensitivity to temporal quality variations.

\section{Loss Functions}
We adopted a composite loss function~\cite{wen2021strong} to enhance accuracy and ranking consistency simultaneously. The first part of the loss function is the precision loss $\mathcal{L}_{\text{p}}$, which quantifies the mean absolute difference between the predicted regression score $\hat{y}\_i$ and the ground truth (MOS) $y_i$. It is defined as:
\begin{equation}
\label{eqn:a04}
\mathcal{L}_{\text{p}} = \frac{1}{N} \sum_{i=1}^{N} \left| \hat{y}_i - y_i \right|.
\end{equation}
The second is the ranking loss $\mathcal{L}_{\text{r}}$, which preserves ordinal consistency by comparing the relative rankings and computing the pairwise differences between sample pairs. It is defined as:
\begin{equation}
\label{eqn:a05}
\mathcal{L}_{\text{r}} = \frac{1}{N^2} \sum_{i=1}^{N} \sum_{j=1}^{N} \max \left( 0, |y_i - y_j| - s_{ij} \cdot (\hat{y}_i - \hat{y}_j) \right),
\end{equation}
where $N$ is the number of videos, and $i$ and $j$ represent the indices of video samples. The sign weight $s_{ij}$ represents the ranking relationship inferred from the ground truth scores and is defined as $s_{ij} = 1$ if $y_i \ge y_j$, and $s_{ij} = -1$ if $y_i < y_j$. 
The composite loss $\mathcal{L}_c$ is formulated as a weighted combination of the two components:
$
\mathcal{L}_c = \lambda_1 \cdot \mathcal{L}_{\text{p}} + \lambda_2 \cdot \mathcal{L}_{\text{r}}.
$
Here, $\lambda \in \mathbb{R}$ balances the two loss terms: $\lambda_1$ improves prediction accuracy, while $\lambda_2$ enhances ordinal consistency. The loss is a weighted combination with $\lambda_1 = 0.6$ and $\lambda_2 = 1$. 

\section{Quality Scoring On Different Dimensions}
\begin{figure*}[htbp]
  \centering
    \includegraphics[width=\linewidth]{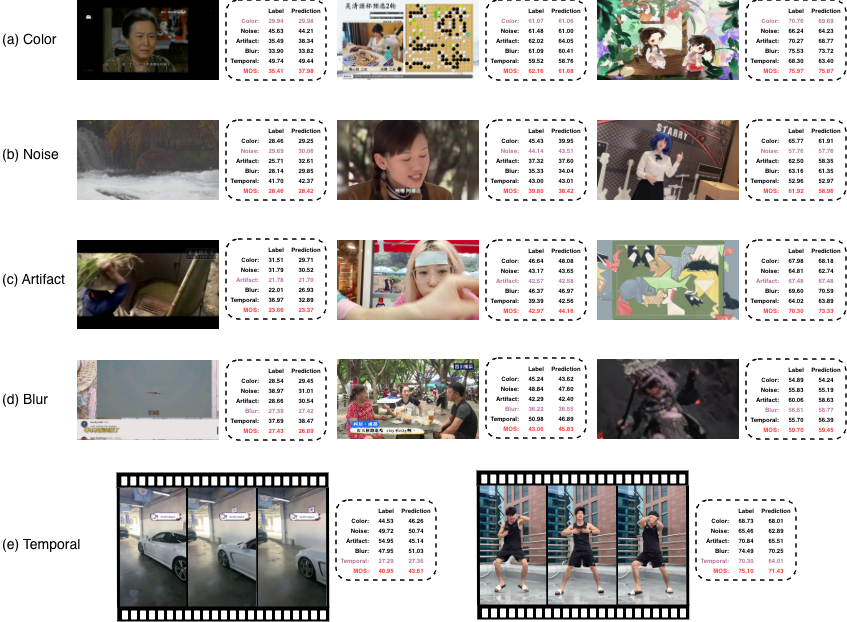}
  \caption{Predicted scores of CAMP-VQA and ground-truth MOS of FineVD~\cite{duan2025finevq} across different quality dimensions}
  \label{fig: MosPredict}
\end{figure*}
Building on our quality-aware captioning, we examined CAMP-VQA’s impact on video quality scoring across different dimensions of the FineVD~\cite{duan2025finevq} dataset. Example videos with predicted CAMP-VQA scores and ground-truth FineVD MOS are shown in Figs.~\ref{fig: MosPredict}.

\end{document}